\documentclass[a4paper,11pt]{article}
\usepackage{pos}
\usepackage{hyperref}
\usepackage{graphicx}
\usepackage[percent]{overpic}
\usepackage{graphbox}

\title{Cross Section Measurements and Searches for New Structures in $e^+e^-$ Annihilations at BESIII}
\author*[a,b]{Hao Liu}
 \onbehalf{on behalf of  BESIII Collaboration}

\affiliation[a]{School of Physical Science and Technology, Lanzhou University, Lanzhou 730000, People's Republic of China}

\affiliation[b]{
Lanzhou Center for Theoretical Physics,
Key Laboratory of Theoretical Physics of Gansu Province,
Key Laboratory of Quantum Theory and Applications of MoE,
Gansu Provincial Research Center for Basic Disciplines of Quantum Physics,
Lanzhou University, Lanzhou 730000, People's Republic of China
}

\emailAdd{lhao2023@lzu.edu.cn}
\abstract{
The BESIII collaboration has collected large data samples in the center-of-mass energy range from 1.84 to 4.95 GeV. These data provide a low-background environment and wide energy range for studying the cross sections and the spectroscopy of charmonia. In this article, we review the recent achievements by the BESIII collaboration, including for searching new structures of $XYZ$ particles and the hyperonic production over open-charm threshold.
}

\FullConference{
The 43rd International Conference on High Energy Physics (ICHEP2026)
}

\begin{document}
\maketitle
\section{Introduction}
The discovery of $J/\psi$ in 1974 by the Samuel C.C.~Ting group~\cite{E598:1974sol} and the B.~Richter group~\cite{SLAC-SP-017:1974ind} marked the birth of the charm quark in the Standard Model.
Since then, charmonium spectroscopy has served as a testing ground for quantum chromodynamics (QCD).
Below the open-charm threshold, the spectrum is well described by potential models and the constituent quark model, as cataloged by the Particle Data Group (PDG)~\cite{ParticleDataGroup:2026mpi}.
Above the $D\bar{D}$ threshold, however, some of the states could not simply be regarded as traditional charmonium~\cite{Wang:2025dur}, namely as $XYZ$ states. 
Theoretically, these states have been interpreted as compact tetraquarks, hadronic molecules, hybrid mesons, or kinematic effects~\cite{Brambilla:2019esw,Guo:2017jvc,Chen:2016qju}. The BESIII experiment~\cite{BESIII:2009fln}, operating at the BEPCII $e^+e^-$ collider~\cite{Yu:IPAC2016-TUYA01} in the center-of-mass energy range from $1.84$ to $4.95$~GeV, has played a leading role in the study of $XYZ$ states.
Since the start of data taking in 2008, BESIII has made remarkable achievements in $XYZ$ spectroscopy, including the discoveries of $X$~\cite{BESIII:2022kqy,BESIII:2024ftl,BESIII:2025xju}, $Y$~\cite{BESIII:2017ysx,BESIII:2017ysy,BESIII:2022gzh,BESIII:2023hqy,BESIII:2023vvt}, $Z$~\cite{BESIII:2013mio,BESIII:2013wzq,BESIII:2014ngc,BESIII:2015cjv,BESIII:2021gdt} and pseudo-tensor $\psi_2$~\cite{BESIII:2015psi2} states.
With the world's largest data samples in this energy region~\cite{BESIII:2015qfd,BESIII:2022ulv,Ablikim:2019hff,Lu:2020imt,Zhang:2022bdc}, BESIII has also performed precise measurements of the cross sections of $e^+e^-$ annihilations for various hadronic pair production processes~\cite{BESIII:2019cuv,BESIII:2023rse,BESIII:2024umc,BESIII:2024ues,BESIII:2024ths,BESIII:2024gql}, enabling the study of resonance line shapes and the search for new vector states via the initial-state radiation (ISR) technique~\cite{BESIII:2017cpu}.

In this article, we review recent highlights from the BESIII collaboration in the study of $XYZ$ states and hyperonic production above the open-charm threshold.
\begin{figure} [!htp]
    \centering
\includegraphics[width=0.95\columnwidth,align=c]{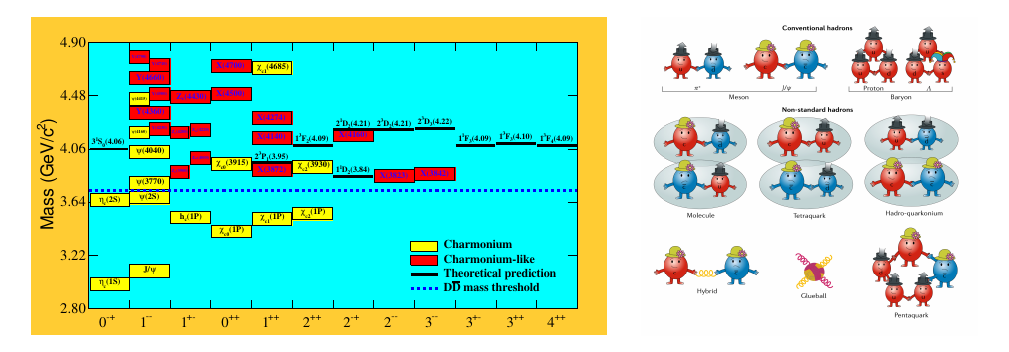}
    \caption{The charmonia(-like) family (left)~\cite{Wang:2025dur} and models of hadrons (right)~\cite{Yuan:2019zfo}.}
    \label{fig:enter-label}
\end{figure}
\section{Highlights on BESIII}
\subsection{Mesonic Production}
\begin{figure}[!htp]
    \centering
    \begin{overpic}[width=0.43\columnwidth]{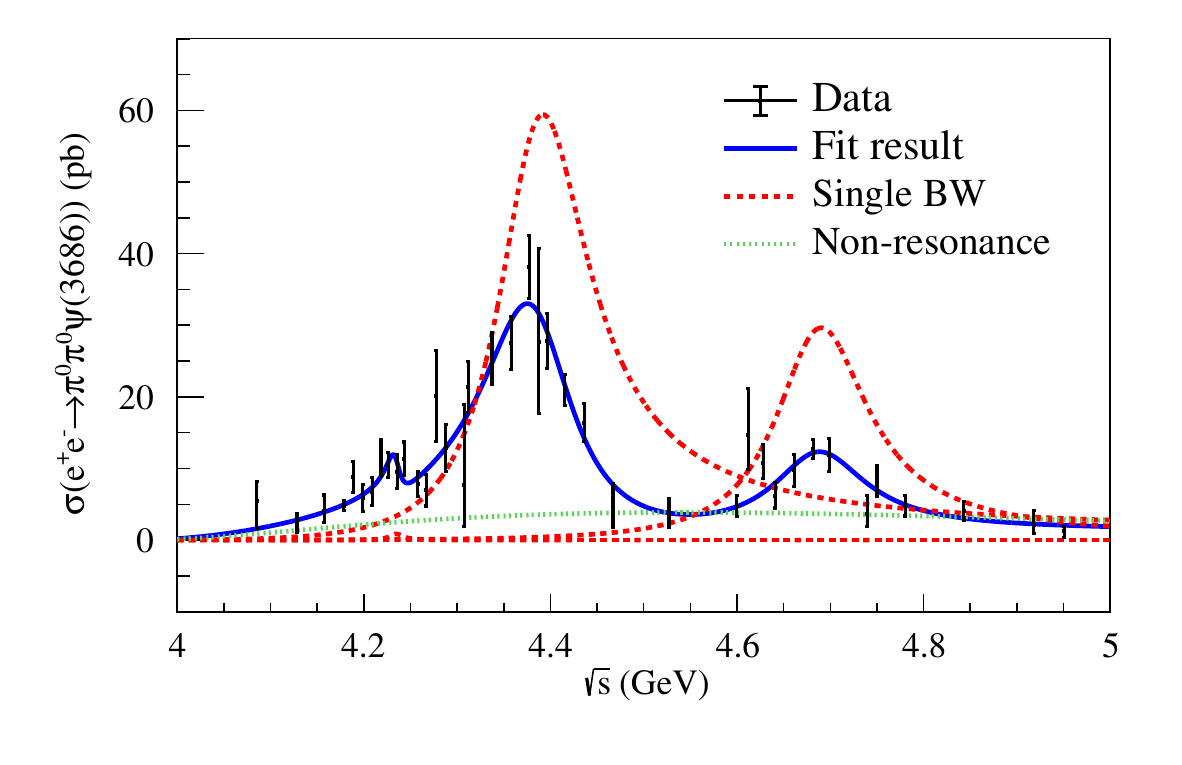}
    \put(25,30){\color{red}\tiny$\psi(4230)$}
    \put(40,57){\color{red}\tiny$\psi(4390)$}
    \put(64,39){\color{red}\tiny$\psi(4660)$}
    \end{overpic}
    \caption{One solution for the fit to the cross section curve of $e^+e^- \to \pi^0 \pi^0 \psi(3686)$~\cite{BESIII:2026qhe}.}
    \label{fig:2pi02S}
\end{figure}

\begin{figure}[!htp]
    \centering
    \includegraphics[width=0.87\columnwidth]{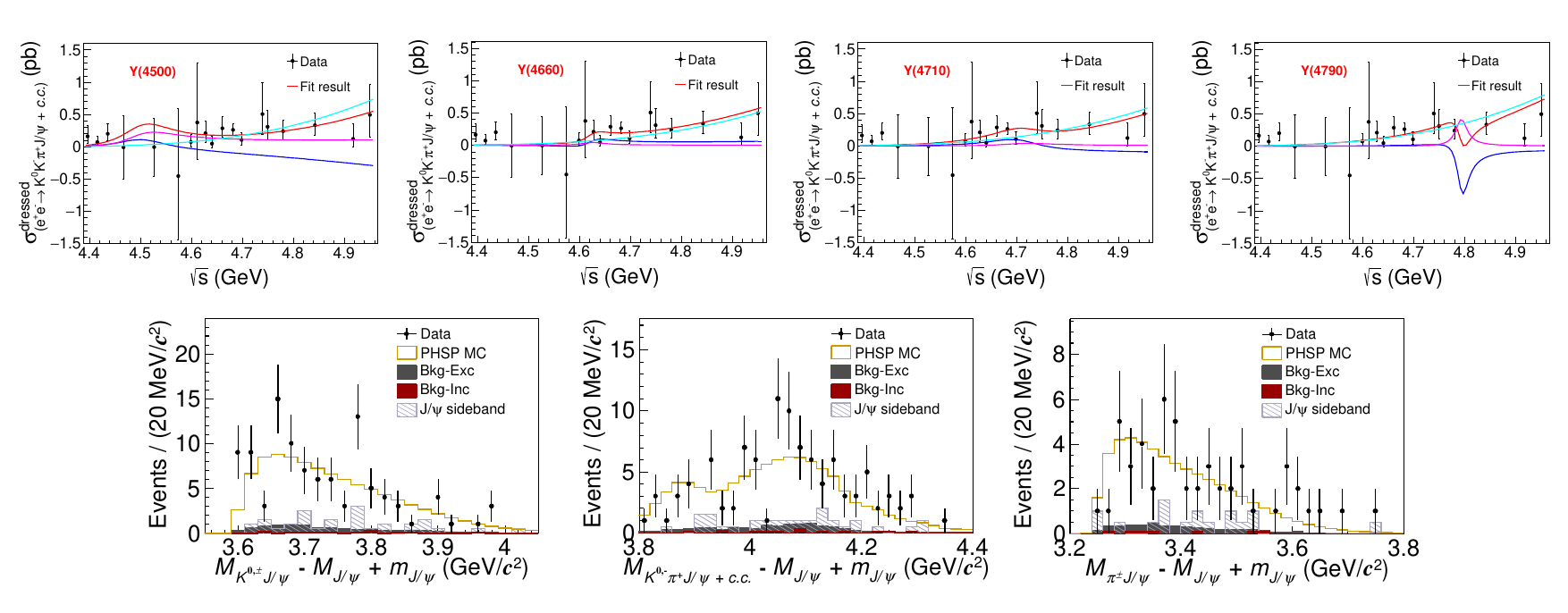}
    \caption{Invariant mass spectra for potential $Z_{c(s)}$ candidate in $e^+ e^- \to K^0 K^+ \pi^- J/\psi$ study~\cite{BESIII:2025fvc}.}
    \label{fig:piKpsi}
\end{figure}

\begin{figure}[!htp]
    \centering
    \begin{overpic}[width=0.83\columnwidth]{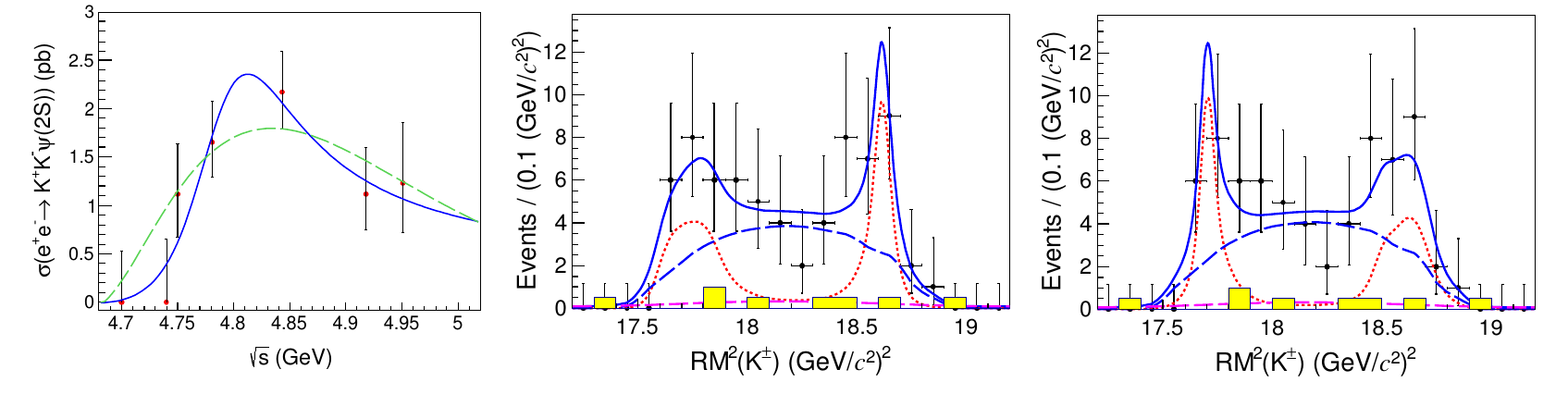}
    \put(27,22){(a)}\put(60.6,22){(b)}\put(94.2,22){(c)}
    \end{overpic}
    \caption{Fit to the cross section (a) and the $Z_{cs}$ candidate in higher (b) and lower (c) position in$e^+ e^- \to K^+ K^- \psi(3686)$~\cite{BESIII:2024vjf}.}
    \label{fig:KK2S}
\end{figure}

The BESIII collaboration has studied the cross section of $e^+ e^- \to \pi^0 \pi^0 \psi(3686)$~\cite{BESIII:2026qhe} aimed at potential $Y$ resonances. In this study, which extends the isospin-conjugate channel previously reported by BESIII, $\psi(4390)$ (\textit{aka} $\psi(4360)$) and $\psi(4660)$ are observed with significances of $11.4\sigma$ and $5.7\sigma$, respectively, while $\psi(4230)$ contributes with only $1.1\sigma$. Fig.~\ref{fig:2pi02S} shows the fit curve of cross sections, which draws the same conclusion as the charged channel. 
For $e^+ e^- \to K^0 K^+ \pi^- J/\psi$~\cite{BESIII:2025fvc} process, the potential $Y$ resonances of $Y(4500)$, $Y(4660)$, $Y(4710)$ and $Y(4790)$ are tested independently, but all of them are not touched with $3\sigma$; additionally, the distribution of $K^{0,\pm}J/\psi$ and $\pi^{-}J/\psi$ mass spectra are also checked, as shown in Fig.~\ref{fig:piKpsi}, but no $Z_{c(s)}$ candidate is found. 
Another study targeting for searching both $Y(4710)$ and $Z_{cs}(4220)$ is performed in the $e^+ e^- \to K^+ K^- \psi(3686)$~\cite{BESIII:2024vjf} study, several fit models are tested, but as shown in Fig~\ref{fig:KK2S}, no candidate is found either.
\subsection{Hyperonic Production}
Recently BESIII has published several studies in hyperonic productions, especially in $e^+ e^- \to \Sigma^- \bar{\Sigma}^+$~\cite{BESIII:2026hie} and $K^+ \Xi^0 \bar{\Sigma}^-$~\cite{BESIII:2026oyx}, both of the processes have confirmed the $\psi(3770)$ resonance with the significances over $5\sigma$. Furthermore, together with the isospin-conjugate channels from previous studies, a systematic comparison can be performed for the triplet of $\Sigma$ hyperon, a detailed estimation of the effective form factor $G_{\rm eff}$~\cite{Faldt:2017kgy} has been completed, furnishing the last piece of the $\Sigma$ triplet. For the other hyperonic studies, like $e^+ e^- \to \Xi(1530)^0 \bar{\Xi}^0$~\cite{BESIII:2026qyk}, $e^+ e^- \to \Omega \bar{\Omega}$~\cite{BESIII:2025fph,Zhang:2025qmo} and $e^+ e^- \to p K^- K^- \bar{\Xi}$~\cite{BESIII:2025lbj}, due to the limited statistics, all assumed resonances, from $\psi(3770)$ to $\psi(4660)$ or $\psi(4710)$, do not reach a significance of $3\sigma$, so only the upper limits at the confidence level of $90\%$ are performed.
\begin{figure}[!htp]
    \centering
    \begin{overpic}[width=0.95\columnwidth,align=c]{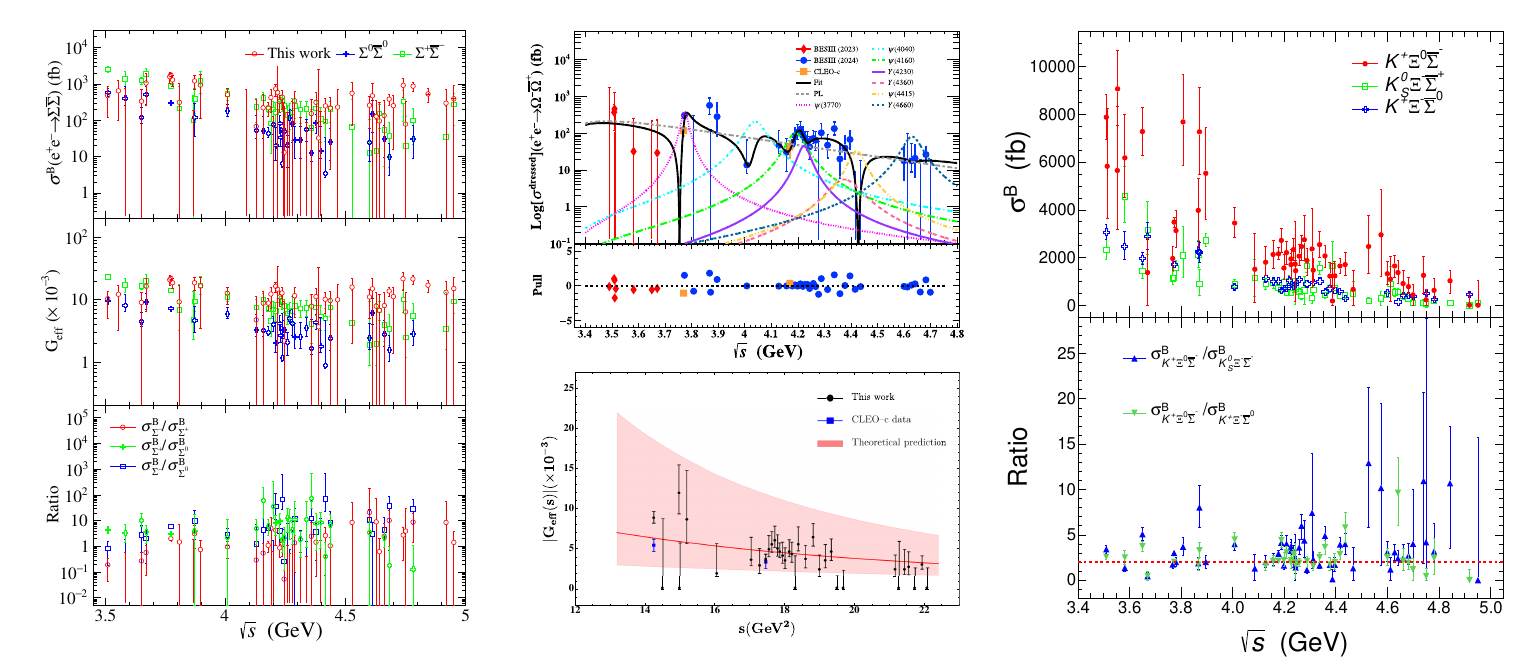}
    \put(26.5,14.5){(a)}\put(59,17){(b)}\put(94,20){(c)}\end{overpic}
    \caption{Study of $e^+ e^- \to \Sigma^- \bar{\Sigma}^+$ in cross section (ratio) and $G_{\rm eff}$ (a)~\cite{BESIII:2026hie}, study of $e^+ e^- \to \Sigma^- \bar{\Sigma}^+$ in cross section and its ratio (b)~\cite{BESIII:2026oyx} and study of $e^+ e^- \to \Omega^- \bar{\Omega}^+$ in cross section and $G_{\rm eff}$ (c)~\cite{BESIII:2025fph,Zhang:2025qmo}.}
    \label{fig:enter-label}
\end{figure}

\section{Summary}
In summary, owing to the wide energy range as well as the large data sample, the BESIII collaboration has enriched the cross section study and $XYZ$ spectra with high precision accepted in PDG~\cite{ParticleDataGroup:2026mpi}, these experimental results could help us to deepen our understanding of QCD behavior. 
Furthermore, based on the interactions and decays of charmonia, BESIII is also a good platform for the further analysis with fine structure, such as the PWA on charmonia.
Recently, the BESIII Collaboration has decided to update the hardware and continue data collection in the following years, and several interesting analyses are now ongoing as well, more exciting results are expected in the future.

\iffalse
\section*{Acknowledgement}
This work is supported in part by
the Fundamental Research Funds for the Central Universities under Contracts Nos. lzujbky-2025-ytA05, lzujbky-2025-it06, lzujbky-2024-jdzx06;
National Natural Science Foundation of China under Contracts No. 12247101;
the Natural Science Foundation of Gansu Province under Contracts Nos. 22JR5RA389, 25JRRA799;
the 111 Project under Grant No. B20063.
\fi

\end{document}